  \documentstyle[12pt]{article}
  
\newcommand{\be}{\begin{equation}}
\newcommand{\ee}{\end{equation}}
\newcommand{\bea}{\begin{eqnarray}}
\newcommand{\eea}{\end{eqnarray}}
\newcommand{\beann}{\begin{eqnarray*}}
\newcommand{\eeann}{\end{eqnarray*}}
\newcommand{\beasn}{\begin{sneqnarray}}
\newcommand{\eeasn}{\end{sneqnarray}}
\newcommand{\bref}[1]{(\ref{#1})}

\newcommand{\ep}{\epsilon}
\newcommand{\vep}{\varepsilon}
\newcommand{\T}{\theta}
\newcommand{\vp}{\varphi}
\newcommand{\A}{\alpha} \newcommand{\B}{\beta}
\newcommand{\G}{\gamma} 

\def\pa{\partial}  
\def\sqr#1#2{{\vcenter{\hrule height.#2pt \hbox{\vrule width.#2pt
height#1pt \kern#1pt \vrule width.#2pt} \hrule height.#2pt}}}

\def\IR{{I\kern-0.25em R}}

\newcommand{\lder}[2]{\frac{\partial_l#1}{\partial #2}}
\newcommand{\rder}[2]{\frac{\partial_r#1}{\partial #2}}

\newcommand{\C}[1]{{\cal #1}}

\newcommand{\NPB}[3]{{\sl Nucl. Phys.} {\bf B#1} (19#2)  {#3}}
\newcommand{\PRD}[3]{{\sl Phys. Rev.} {\bf D#1} (19#2)   {#3}}
\newcommand{\PLB}[3]{{\sl Phys. Lett.} {\bf #1B} (19#2)  {#3}}

\newcommand{\PRep}[3]{{\sl Phys. Rep.} {\bf #1} (19#2)  {#3}}

\catcode`@=11
\def\dif{{\rm d}}
\def\deriv{\@ifnextchar[{\@deriv}{\@deriv[]}}
   \def\@deriv[#1]#2#3{\mathchoice%
{{\dif^{#1}#2\over\dif{#3}^{#1}}}{{\dif^{#1}#2/\dif{#3}^{#1}}}%
{{\dif^{#1}#2\over\dif{#3}^{#1}}}{{\dif^{#1}#2/\dif{#3}^{#1}}}}

\def\presup#1{{}^{#1}\kern-.15em\relax}      
\def\presub#1{{}_{#1}\kern-.12em\relax}      

\def\secteqno{\@addtoreset{equation}{section}%
\def\theequation{\thesection.\arabic{equation}}}
\def\endsecteqno{\def\theequation{\@ifundefined{chapter}%
{\arabic{equation}}{\thechapter.\arabic{equation}}}}
\newcounter{subequation}
\def\thesubequation{\alph{subequation}}
\def\sneqnarray{\stepcounter{equation}\let\@currentlabel=\theequation
\setcounter{subequation}{1}
\def\@eqnnum{{\rm (\theequation\thesubequation)}}
\global\@eqcnt\z@\tabskip\@centering\let\\=\@eqncr\let\@@eqncr=\@@sneqncr
$$\halign to \displaywidth\bgroup\@eqnsel\hskip\@centering
 $\displaystyle\tabskip\z@{##}$&\global\@eqcnt\@ne
 \hskip 2\arraycolsep \hfil${##}$\hfil
&\global\@eqcnt\tw@ \hskip 2\arraycolsep $\displaystyle\tabskip\z@{##}$\hfil
  \tabskip\@centering&\llap{##}\tabskip\z@\cr}
\def\endsneqnarray{\@@sneqncr\egroup $$\global\@ignoretrue}
\def\@@sneqncr{\let\@tempa\relax
   \ifcase\@eqcnt \def\@tempa{& & &}\or \def\@tempa{& &}
   \else \def\@tempa{&}\fi
     \@tempa \if@eqnsw\@eqnnum\stepcounter{subequation}\fi
     \global\@eqnswtrue\global\@eqcnt\z@\cr}
\def\nobiblabels{\def\@lbibitem[##1]##2{\@bibitem{##2}}}
\catcode`@=12

\secteqno

\title{{\bf One Loop Anomalies and Wess-Zumino
 Terms for General Gauge Theories}}
\author{{\sc J. Gomis}$^\diamondsuit$,
        {\sc K.Kamimura}$^\clubsuit$,
        {\sc J.M. Pons}$^{\heartsuit,\spadesuit}$
        {\sc and F. Zamora}$^\spadesuit$\\
        \llap{$^\diamondsuit$}%
        \small{\it{Research Institute for Mathematical Sciences}}\\
        \small{\it Kyoto University,}
        \small{\it{Kyoto 606-01}}\\
        \small{\it{JAPAN}}\\
        \llap{$^\clubsuit$}%
        \small{\it{Department of Physics, Toho University}}\\
        \small{\it{Funabashi}}\\
        \small{\it{274 JAPAN}}\\
        \llap{$^\heartsuit$}%
        \small{\it{Center for Relativity}}\\
        \small{\it{Department of Physics, University of Texas at Austin}},
        \small{\it{Austin, 78712 Texas}}\\
        \small{\it{USA}}\\
       \llap{$^\spadesuit$}%
        \small{\it{Departament d'Estructura i Constituents
               de la Mat\`eria}}\\
        \small{\it{Universitat de Barcelona and }}\\
        \small{\it{Institut de F\'{\i}sica d'Altes Energies}}\\
        \small{\it{Diagonal, 647}}\\
        \small{\it{E-08028 BARCELONA}}\\
{\it e-mails:} \small{gomis@kurims.kyoto-u.ac.jp, kamimura@jpnyitp,}\\
                       \small{pons@rita.ecm.ub.es, zamora@ecm.ub.es}}

\date{}

\begin{document}

\maketitle

\thispagestyle{empty}

\begin{abstract}
One loop anomalies and their dependence on antifields for general gauge
theories are investigated within a Pauli-Villars regularization scheme.
For on-shell theories {\it i.e.}, with open algebras or  on-shell
reducible theories, the antifield dependence is  cohomologically non
trivial. The associated  Wess-Zumino term depends also on antifields.
In the classical basis the antifield independent part of the WZ term
is expressed in terms of the anomaly and finite gauge transformations
by introducing gauge degrees of freedom as the extra dynamical variables.
 The complete WZ term is reconstructed from
the antifield independent part.

\end{abstract}
\vskip 10mm

\vfill
\vbox{
\hfill July 1995 (revised Oct. 1995)\null\par
\hfill RIMS-1020\null\par
\hfill TOHO-FP-9552\null\par
\hfill UB-ECM-PF 95/17}\null


 \section{Introduction.}
 \indent

 Most known fundamental theories are gauge theories.
 The deep knowledge of their gauge structure is crucial
 to understand the classical and the quantum natures completely.
 The recent work of strong-weak duality in supersymmetric gauge
theories \cite{sw} suggests that our understanding of the gauge
properties is incomplete and new ideas to understand these issues
seem to be required. In this paper we address a more traditional
aspect of gauge theories. We will study the structure of gauge
anomalies and the form of the WZ term  \cite{wz}
for general gauge theories using the BRST symmetry \cite{brs}\cite{t}.
In particular we will discuss a parametrization of the one loop anomaly
and the associated WZ term using a PV regularization scheme\cite{tnp}.
We will see how the anomaly can be expressed  in terms of PV
regulator and the BRST transformation \cite{gp2}. This form of the
anomaly is very useful in order to study the antifield dependence.
For theories with on-shell structure, {\it i.e.}, with open algebras
or on-shell reducible theories \cite{wh} \cite{bv},
the anomalies will be dependent in a non-trivial way on the sources
(antifields) of the BRST transformation.
 While for closed theories this dependence is cohomologically trivial.
However, using the WZ consistency conditions and cohomological techniques
it is possible to construct for closed theories antifield dependent
candidate anomalies
 \cite{b}. They cannot appear in any regularized field theory
 calculation.
The regularization procedure selects a subset of the ghost number one
nontrivial cocycles.

The presence of gauge anomalies implies that some classical gauge
 degrees of freedom become dynamical (propagating) at quantum level.
 The WZ term can be written using these new degrees of freedom.
 In general the WZ term also depends on the sources of the BRST
 transformation. The antifield independent part of the WZ term
 in the classical basis of the fields and the antifields
 can be expressed in terms of the anomaly and the finite gauge
 transformations associated with the on-shell structure
 as in the ordinary case, like
  Yang-Yills theory \cite{z}.

The organization of the paper is as follows: in section 2 we will
introduce some basic concepts and the effective action
regularized by PV scheme up to one loop.
In section 3 we will discuss the structure of the anomaly. In section 4
the WZ term is analyzed. In section 5 we use the non-abelian antisymmetric
tensor field \cite{ft} to illustrate our formalism.
We give some conclusions in the last section.
Some discussions of finite on-shell structures are in Appendix.

 \section{One loop PV regularized effective action}
 \indent

 Let us consider a general gauge theory with classical action
 $ \C{S}_0(\phi)$. We assume its infinitesimal gauge transformations
 $ \delta\phi=R^i_{~\A}(\phi)\epsilon^\A$
 have  a reducible on-shell open algebra structure
 \footnote{To make notation simpler we will
 consider only bosonic fields and bosonic gauge transformations.}.
 Some relations of the gauge structure functions are :
 \be
 \label{openRR}
 {R^j}_{~[\A}(\phi) {R^i}_{\B],j} (\phi) + T^{\G}_{\A \B}(\phi)
 {R^i}_\G (\phi)
 + E^{ij}_{\A \B}(\phi) {\cal S}_{0,j}~=~0,
 \ee
  \be
 R^i_\A(\phi) Z^\A_a(\phi) ~+~ V^{ij}_a(\phi) {\cal S}_{0,j}~=~0
 \label{red-off}
 \ee
and
\be
\label{4-4}
\sum_{P\in{\rm Perm}[\A\B\G]} (-1)^P(~T^\delta_{\B\G,j}~R^j_{~\A}~+
T^\delta_{\mu\G}~T^\mu_{\A\B}~+~Z^\delta_a~F^a_{\A\B\G}~-~
D^{j\delta}_{\A\B\G}~S_{0,j}~)~=~0,
\ee
where $~{\cal S}_{0,j}\approx 0~$ is the classical equation of motion.
 The first one implies that the algebra closes on shell,
 the second one means the transformation is reducible on shell
and the third one is the generalized Jacobi identity.

 The whole set of such relations can be expressed in a compact way
 within the Field-Antifield formalism \cite{bv}
 \footnote{For  reviews see \cite{ht}\cite{gps}\cite{tp}.}.
 We denote all the fields by $\Phi^A$
 (classical fields
 $\phi^i$, $i=1,...,n$; ghosts $c^\A$, $\A=1,...,m_0$; antighosts
 ${\bar c}_\A$; ghost for ghosts $\eta^a$, $a=1,...,m_1$; etc)
 and their corresponding antifields (Afs) by $\Phi^\ast_A$. The proper
 solution $\C{S}(\Phi,\Phi^*)$ of the Classical Master Equation (CME)
 \be
 \label{CME}
 (\C{S},\C{S})=0
 \ee
 admits the local expansion in Afs,
 \be
 \label{ClPS}
 {\cal S}(\Phi,\Phi^\ast)= \C{S}_0(\phi) + \Phi^\ast_A
 \C{S}^A_1(\Phi) + \frac{1}{2} \Phi^\ast_A \Phi^\ast_B
 \C{S}^{BA}_2(\Phi) + \, ... \, .
 \ee
 The CME (\ref{CME}) encodes all the relations among these structure
functions  such as \bref{openRR}-\bref{4-4} and is determining the
complete classical gauge structure \cite{bv85}\cite{fh}.
 The terms $\C{O}({\Phi^\ast}^2)$ in the proper solution appear
 for theories with open or  reducible on-shell algebras.
 This basis reproducing  all the classical gauge  structure is called
classical basis(ClB).  The BRST transformation in the space of fields
and antifields is generated by
 $\delta\cdot=(~\cdot~, \C{S})$ and it is nilpotent off shell,
 \be
 \delta^2~=~0.
\label{brstcoho}
\ee
 Even for fields this transformation gives terms depending
 in general on Afs,
 \be
 \delta \Phi^A = (\Phi^A,\C{S}) = \C{S}^A_1(\Phi) +
\Phi^\ast_B \C{S}^{BA}_2(\Phi) + \, ...~.
 \ee

 To perform perturbative calculations the basis is changed from the ClB
to the gauge fixed basis (GFxB).
This change   is implemented by an antibracket canonical transformation
\cite{vt} in general.
Often discussed are those generated by gauge fixing fermions $\Psi(\Phi)$.
Under such canonical transformation the fields are unchanged  while
Afs $\Phi^*_A$ are transformed to new Afs  $K_A$ in GFxB as
 \be
 \label{ct to gfb}
 \Phi^\ast_A = K_A + \frac{\delta
 \Psi(\Phi)}{ \delta \Phi^A} \, .
 \ee
 In GFxB the proper solution reads
 \bea
 \hat{\C{S}}(\Phi,K)&& =
 \left[ \C{S}_0 + \frac{\delta \Psi}{\delta \Phi^A} \C{S}^A_1
 + \frac{1}{2} \frac{\delta \Psi}{\delta \Phi^A} \frac{\delta
 \Psi}{\delta \Phi^B} \C{S}^{BA}_2 +\, ... \, \right]
 \nonumber
 \\
 &&~~~~ + \, K_A \left[ \C{S}^A_1 + \frac{\delta \Psi}
 {\delta \Phi^B} \C{S}^{BA}_2 + \frac{1}{2} \frac{\delta \Psi}{\delta
 \Phi^B} \frac{\delta \Psi}{\delta \Phi^C} \C{S}^{CBA}_3 + \, ... \,
 \right] + \, ...
 \nonumber
 \\
 \label{GFxPS}
 && =: \C{S}_\Sigma(\Phi) + K_A \left[\delta_\Sigma \Phi^A \right] + \,
 ...\,.
 \eea
 The Afs independent part of the proper solution,
 $\C{S}_\Sigma(\Phi)$, has no more gauge invariance and
 gives well defined propagators. It is invariant
 under the Afs independent gauge fixed BRST operator
 $\delta_\Sigma$ that acts
 on local functionals $F(\Phi)$ of the fields by
 \be
 \delta_\Sigma~ \cdot~ =~ (~\cdot~ , \hat\C{S})|_{K=0}.
 \ee
 One can verify that $\delta_\Sigma$ is nilpotent on-shell,
 \be
 \label{WN}
 \delta_\Sigma^2 \simeq 0 \,.
 \ee
 Here the weak equality ``$\simeq$'' means that it holds
 up to gauge fixed equations of motion $\C{S}_{\Sigma,A}\simeq 0$.
 Nilpotent operators $\delta$ and $\delta_\Sigma$
 are associated with the antibracket cohomology and the gauge fixed weak
 cohomology respectively. Their relation has been studied in
 \cite{hf}.

 The quantum aspects of the BRST formalism are most suitably studied
 in terms of the effective action $\Gamma$ which is obtained from the
 (connected part of) the generating functional by a Legendre
 transformation with respect to the sources $J_A$.
 We will consider a PV regularization at one loop level.
 The generating functional is

 \be
   Z_{\rm reg}(J,K)=\int{\cal D} \Phi {\cal D} \chi
   \exp\left\{\frac{i}{\hbar} \left[
\hat{\C{S}}(\Phi,K)+\hbar \hat\C{M}_1(\Phi, K)
   + S_{\rm PV}(\chi,\chi^*=0;\Phi, K)+J_A\Phi^A\right]\right\},
 \label{reg gen func}
 \ee
 where  $\chi^A$ are
 the PV fields.
 Each PV field $\chi^A$ comes with its associated antifield
 $\chi^*_A$, and they can collectively be denoted as
 $w^a=\{\chi^A, \chi^*_A\}$, $a=1,\ldots,2N$. PV antifields
 $\chi^*_A$ have no physical significance and are put to zero at the end.
 The local counter term
 $ \hat\C{M}_1(\Phi, K)$ should guarantee the finiteness of theory while
preserve the BRST structure at quantum level if it is possible.

 The PV action $S_{\rm PV}$ is determined from two requirements:
 i)\quad massless propagators and couplings for PV fields should coincide
 with those of their partners and
ii)\quad BRST transformations for PV fields should be such that the
massless part of the PV action, $S_{\rm PV}^{(0)}$, and the measure in
\bref{reg gen func} be BRST invariant up
to one loop. A suitable prescription for $S_{\rm PV}$
is \cite{tnp} \cite{tp}
 \be
   S_{\rm PV}= S^{(0)}_{\rm PV}+S_M=
     \frac12 w^a S_{ab} w^b
     -\frac12 M \chi^A T_{AB} \chi^B,
 \label{orig pv action}
 \ee
 with the mass matrix $T_{AB}$ being invertible but otherwise arbitrary
 and $S_{ab}$ is defined by
 \be
     S_{ab}=\left(\lder{}{{z}^a}\rder{}{{z}^b}\hat{S}(\Phi, K)\right),
 \label{s der}
 \ee
 where $z^a=\{\Phi^A, K_A \}$ and $~l,~r~$ stand for left and right
derivatives.

  Application of the semiclassical approximation to \bref{reg gen func}
 yields the regularized effective action up to one loop \cite{gp2}
 \be
 \label{effectA}
    \Gamma(\Phi, K)= \hat{\C{S}}(\Phi,K)+\hbar \hat\C{M}_1(\Phi,K)+
\frac{i\hbar}2{\rm Tr~Ln}\left[\frac{(T{\hat\C{R}})}{(T{\hat\C{R}})-(TM)}
\right]  \equiv
    \hat{\C{S}}(\Phi,K)+\hbar \Gamma_1(\Phi,K).
 \label{gamma1}
 \ee
 Here Tr stands for the supertrace and
 $(T{\hat\C{R}})_{AB}$ is defined from \bref{s der} as
 \be
\label{regulator}
    (T{\hat\C{R}})_{AB}=\left(\frac{\partial_l}{\partial\Phi^A}
    \frac{\partial_r}{\partial\Phi^B} \hat{S}(\Phi, K)\right).
 \label{orig pv kin}
 \ee
 The denominator of the ~"Tr~Ln"~ term in \bref{effectA}
 is the contribution from the integrations over the PV fields
 and the numerator is the one from the quantum fluctuation of $\Phi$.
The PV regulator ${\hat\C{R}}$ is obtained from \bref{regulator} and the
form of the mass matrix $T_{AB}$ in the PV action \bref{orig pv action}.

 \section{Antifield Dependence of the Anomaly }
\indent

 Once we obtain the effective action we can calculate the anomalous
  Slavnov-Taylor identity \cite{s}\cite{l}\cite{Zinn}
\be
(\Gamma,\Gamma)=-i\hbar \hat{\cal A}_1\cdot\Gamma.
\ee
 Up to one loop we have

 \be
 \label{ST}
 (\Gamma_1,\hat \C{S})=-i\hat{\cal A}_1
 \ee
 where
 \be
 \label{ano1}
 \hat{\cal A}_1=:\Delta \hat{\C{S}}+i(\hat\C{M}_1,\hat{\C{S}})
 \ee
 is the potential anomaly. More explicitly  using \bref{effectA},
 \bref{ST} and  \bref{ano1} we obtain \cite{gp2}\cite{vp}

 \be
 \label{PV}
 \Delta \hat\C{S}(\Phi,K) =
  \delta \left\{ -\frac{1}{2} {\rm Tr Ln}
\left[ \frac{{\hat\C{R}}(\Phi,K)}{{\hat\C{R}}(\Phi,K)-{M}}\right]
 \right\} \, .
 \ee

$\hat{\cal A}_1 $ measures the BRST non invariance of the effective
action and $\Delta \hat{\C{S}}$ reflects the non invariance of
the path integral measure
 \be
 \C{D}\Phi \ \longrightarrow \ \C{D}\Phi(1+\Delta \hat\C{S})
 \ee
under the classical BRST transformations
$\delta \Phi^A = (\Phi^A,\hat\C{S})$.
The form of (\ref{PV}) shows $\Delta \hat\C{S}(\Phi,K)$ is satisfying
the Wess-Zumino consistency conditions (WZCC),
$\delta(\Delta \hat\C{S})= 0$.

    $\Delta \hat{\C{S}}(\Phi,K)$ can be rewritten as
 \be
 \label{delSreg}
 \Delta \hat{\C{S}}(\Phi,K)=
 {\rm Tr}\left[-\frac12({\hat\C{R}}^{-1}\delta {\hat\C{R}})
    \frac{1}{(1-\frac{\hat\C{R}}{M})}\right].
 \label{a 2 quad}
 \ee
This parametrization of the anomaly is very important in order to study
the antifield dependence. Furthermore  this expression shows that
(potentially) anomalous symmetries are directly related with the
transformation properties of the regulator ${\hat\C{R}}$. In particular,
if ${\hat\C{R}}$ is invariant under some subset of symmetries or it
is transformed as  $\delta {\hat\C{R}}=[{\hat\C{R}}, {\C{G}}]$ by
some operator ${\C{G}}$, the Tr of \bref{a 2 quad} leads to a
vanishing result. Notice that the anomaly depends on both fields and
antifields.

In order to study the antifield dependence of the anomaly we should
pass to the ClB\footnote{We acknowledge Mark Henneaux
discussing on this point.}.
 By performing a canonical transformation, the inverse of \bref{ct to gfb},
 we go to the ClB.
 Since the Jacobian of the transformation is 1 the
 expression $\Delta\C{S}(\Phi,\Phi^*)$ in ClB is given by expressing
 $\Delta\hat\C{S}(\Phi,K)$ in terms of the variables $\Phi,\Phi^*$
in ClB,
 \be
 \Delta\C{S}(\Phi,\Phi^*)~\equiv
 \Delta\hat\C{S}(\Phi,\Phi^\ast_A - \frac{\delta \Psi(\Phi)}
{ \delta \Phi^A} ) =( -\frac{1}{2} {\rm Tr Ln}
\left[ \frac{{\C{R}}(\Phi,\Phi^*)}{{\C{R}}(\Phi,\Phi^*)-{M}}\right],
 {\C{S}}(\Phi,\Phi^*))
 \ee
 Here $\C{R}(\Phi,\Phi^*)~=~
 \hat\C{R}(\Phi,\Phi^\ast_A - \frac{\delta \Psi(\Phi)}{ \delta \Phi^A})$
is the regulator in the ClB.

It is also useful to consider
 first the following BRST operator in classical basis. It
 acts on functions depending only on fields as
 \be
 \delta_0~ \cdot~  =~(~\cdot~, \C{S})|_{\Phi^*=0}.
 \ee
 It is nilpotent on shell, {\it i.e.} on the classical equations of
motion; $~{\cal S}_{0,j}\approx 0~$; ~
\be
\delta_0^2\approx 0.
\ee
 Next by expanding  ${\cal R}(\Phi,\Phi^*)$ with respect to $\Phi^*$ as

 \be
 {\C{R}}(\Phi,\Phi^*) = \C{R}(\Phi) + \Phi^*_A \C{R}^A (\Phi) +
 \frac{1}{2!} \Phi^*_A \Phi^*_B \C{R}^{BA}(\Phi) + \, ... \, .
 \ee
we obtain
 \bea
 \nonumber
 \Delta \C{S}(\Phi,\Phi^*) && = \delta_0
 \left\{ -\frac{1}{2} {\rm Tr Ln}
\left[\frac{\C{R}(\Phi)}{\C{R}(\Phi)- {M}}\right] \right\}
 \\
 \nonumber
 && -
 {\rm Tr} \left\{-\frac{1}{2} \C{R}(\Phi)^{-1}
 \C{R}^i(\Phi)\left[\frac{1}{1-
 \frac{\C{R}(\Phi)}{{M}}}\right] \right\}
\,\frac{\delta_l\C{S}_0}{\delta\Phi^i}
 + \, \C{O}(\Phi^*)
 \\
 &&=\delta_0(-\frac{1}{2} {\rm Tr Ln}
\left[ \frac{{\cal R}(\Phi)}{{\cal R}(\Phi)-{M}}\right])
 -\, P^i(\phi) \frac{\delta_l S_0}{\delta\phi^i}  + \, \C{O}(\Phi^*) \, .
 \label{AfIPV}
 \eea
 Using the fact that the non-minimal sector is cohomologically trivial
in ClB it can be written  as
 $$
 \Delta\C{S}(\Phi,\Phi^*)~=~\C{A}_\alpha(\phi)c^\alpha+
(P^i\phi^*_i+N,S)+ \, \C{O}(\Phi^*) \,,
 $$
 where
 \be
\label{classical}
 \C{A}_\alpha(\phi)c^\alpha=
 \delta_0( -\frac{1}{2} {\rm Tr Ln}
\left[ \frac{{\cal R}(\phi)}{{\cal R}(\phi)-{M}}\right])
 \ee
 and ${\cal R}(\phi)$ is the part of ${\cal R}(\Phi)$
 depending only on the classical fields.
 $N$ is the local counter term that reproduces the
   contribution from non-minimal fields.
 Notice that $\delta_0 \C{A}_\alpha(\phi)c^\alpha \approx 0$.
 $ \Delta\C{S}(\Phi,\Phi^*)$ can be reconstructed
 once the antifield independent part is known.
 The general procedure was first discussed in
\cite{h} and later in \cite{vp}.
 Up to cohomologically trivial terms we have
\be
 \Delta\C{S}(\Phi,\Phi^*)~=~\C{A}_\alpha(\phi)c^\alpha+
\Phi_A^*\C{A}^A(\Phi)+
 \frac{1}{2!} \Phi^*_A \Phi^*_B \C{A}^{BA}(\Phi) + \, ... \, .
\ee

  Only for theories with on shell algebras
 the anomaly will depend in a nontrivial way on the antifields.
 For theories with closed algebras and reducible off-shell algebras,
their actions are linear in antifields and thus
$\delta=\delta_0$, $\C{A}_\alpha(\phi)c^\alpha$ is off shell BRST invariant.
For these theories the antifield dependent part of the anomaly is
trivial. However, using the WZ consistency conditions and cohomological
techniques it is possible to construct for closed theories antifield
dependent candidate anomalies  \cite{b}; these candidates cannot appear
in any regularized field theory calculation.
The regularization procedure selects a subset of the ghost number one
nontrivial cocycles.

The antifield independent part of the anomaly in GFxB is obtained from the
one in ClB by the canonical transformation
 \bref{ct to gfb} as
\be
 \Delta\hat{\C{S}}^{GFxB}(\Phi)~=~\C{A}_\alpha(\phi)c^\alpha+
(\frac{\delta \Psi({\Phi})}{ \delta {\Phi^A}})\C{A}^A(\Phi)+
 \frac{1}{2!} (\frac{\delta \Psi({\Phi})}{ \delta {\Phi^A}})
 (\frac{\delta \Psi({\Phi})}{ \delta {\Phi^B}})
  \C{A}^{BA}(\Phi) + \, ... \, .
\ee

 \section{ The WZ term}
 \indent

 If $\Delta\hat\C{S}(\Phi,K)$ found in GFxB is different from zero
the path integral measure is not BRST invariant. In some cases it
is possible to restore the BRST
invariance by a suitable choice of the local counter term
$\hat\C{M}_1(\Phi,K)$  such that $\hat{\cal A}_1=0$.
In case no local counter term exists we have a genuine
anomalous theory. Physically it means that some classical gauge degrees
of freedom turn to be dynamical at quantum level.
We can consider the gauge parameters
 as  these extra degrees of freedom.
A natural question arises,
when we enlarge the space of fields to $\Phi^A$ and
these extra degrees  of freedom \cite{fs}\cite{gp1},
whether it is possible to find a local counter term
 $ \hat\C{M}_1(\tilde\Phi,K)$ such that its BRST variation
 $\tilde\delta$ in the extended space gives the anomaly,
 \be
\label{delMg}
 \tilde\delta \hat\C{M}_1(\tilde\Phi,K)=i\hat{\cal A}_1(\Phi,K),
 \ee
where $\tilde\Phi$ is a collective notation of $\Phi$ and
the extra variables  corresponding to gauge degrees  of freedom
$\theta^\alpha$ as well as possible redundant gauge
freedoms for reducible theories.
This local counter term becomes the WZ term in the extended formalism.

To find the WZ term we first consider the equation in
ClB which is corresponding to \bref{delMg} in GFxB
 \be
 \tilde\delta \C{M}_1(\tilde\Phi,\Phi^*)=
 i{\cal A}_1(\Phi,\Phi^*).
 \ee
 Due to the cohomological reconstruction procedure in ClB
 it is sufficient to find its antifield independent part
 $\C{M}_1(\phi,\theta^\alpha)$  verifying
  \be
 \label{WZeq AI Cl}
\tilde\delta_0  \C{M}_{1}(\phi,\T) \approx  i\,
 {\C{A}}_{\A}(\phi) c^\A \, ,
 \ee
where ${\cal A}_{\A}(\phi)c^\A = {\cal A}_1(\Phi,\Phi^\ast=0)$
and $\C{M}_{1}(\phi,\T)$ are the Afs independent part of the anomaly
and the WZ term, respectively, in the ClB.
The interesting property of the classical basis is that we can find a
solution of $\C{M}_{1}(\phi,\T)$ depending only on $\phi$ and $\theta$.
A particular non-local solution of \bref{WZeq AI Cl}
have been obtained in \bref{classical}.

We can write the general solution of the homogeneous equation
as an arbitrary function of variable $F^i(\phi,\T)$,
which is the finite transformation of $\phi^i$, because
we can introduce transformation properties of the extra variables
(gauge parameters) such that $\tilde\delta_0 F^i(\phi,\T)\approx 0$.
In fact using a relation obtained from the on-shell composition law
\bref{eqtrB} and
\be
\label{ocl}
0 \approx \tilde\delta_0 F^i(\phi,\T)=\frac{\partial F^i(\phi,\T)}
{\partial \phi^j}~R^j_\A(\phi)~\ep^\A +~
\frac{\partial F^i(\phi,\T)}{\partial \T^\beta}~\delta \T^\A
\ee
we find
\be
\label{ocl1}
\delta \T^\A =
- \tilde\mu^\A_{~\B}(\T, \phi)\ep^\B + Z^\A_a(\T, \phi)\ep^a,
\ee
where
$ \tilde \mu^\B_{~\A}(\T,\phi) :=\left.\frac{\partial\vp^\B(\T',\T;\phi)}
{\partial\T'^\A}\right|_{\T'=0}$ and
$ Z^{\alpha}_{a}(\T, \phi) := \left. \frac{\partial f^\A (\T,\ep,\phi)}
 {\partial \ep^a}\right|_{\ep = 0}$ is a nullvector of
$\frac{\partial F^i(\phi,\T)}{\partial \T^\beta}~$, see \bref{eqtrC}.
 The algebra of transformations of $\phi$ and $\theta$
 remains  to be open and on-shell reducible.
 Therefore we have a new realization of the  on-shell structure.

Now we can write a solution of \bref{WZeq AI Cl}
 as a sum of the particular
solution and the general solution of the homogeneous equation,
 \be
 \label{gensol}
 {\C{M}}_{1}(\phi,\T)={\C{M}}_{1}^{non}(\phi)+G(F(\phi,\theta)),
 \ee
 where ${\C{M}}_{1}^{non}(\phi)$ is the non-local solution
 obtained from \bref{classical}.
 The function $G(F)$ in \bref{gensol} is fixed if we impose a 1-cocycle
 condition for the on shell structure
 \be
 \label{cocycle}
 {\C{M}}_{1}(\phi,\varphi(\T,\T';\phi))~\approx~{\C{M}}_{1}
 (F(\phi,\T),\T')~+~{\C{M}}_{1}(\phi,\T),
 \ee
 we get
 \bea
 \label{cocycle1}
 {\cal M}_{1}(\phi,\T) && \approx \, {\cal M}^{non}_{1}(\phi)
 - {\cal M}^{non}_{1}(F^i(\phi,\T)).
 \eea
 If we write \bref{cocycle1} as a surface integral in a variable $t$ and
 we use the on-shell Lie equations \bref{eqtr} we have
\bea
\nonumber
{\cal M}_{1}(\phi,\T)
&& \approx - \int_0^1 {\dif t} \, \frac{d}{\dif t} {\cal
M}^{ non}_{1}(F(\phi,t\T)) = - \int_0^1 {\dif t} \,
\frac{\pa{\cal M}^{non}_{1}(F(\phi,t\T))}{\pa F^i}
\frac{\pa F^i(\phi,t\T)}{\pa (t\T^\B)} \T^\B
\\
&& \approx - \int_0^1 {\dif t} \, \left.\frac{\pa{\cal M}^{non}_{1}}
{\pa\phi^i}\right|_{\phi=F(\phi,t\T)} R^i_\A(F(\phi,t\T))
\lambda^\A_\B(t\T,\phi) \T^\B \, ,
\label{wz}
\eea
which can be written using \bref{classical} and \bref{WZeq AI Cl}
up to equation of motion as
 \be
 \label{WZ AI Cl}
 {\cal M}_{1}(\phi,\T) = \, -i \int_0^1 {\dif t} \,
 {\cal A}_{\A}(F(\phi,t\T)) \lambda^\A_{~\B}(t\T,\phi) \T^\B.
 \ee
 Notice that this antifield independent part of the WZ term in the
ClB has the same form as that for closed theories and
off-shell reducible theories  \cite{z}\cite{gp1}\cite{gp2}\cite{gpz}.
 Now applying the Afs perturbative method
 we can find the full WZ term
 \be
 \label{WZ AI Cl2}
 {{\cal M}}_1({\tilde \Phi},{ \Phi^\ast}) = { {\cal M}}_{1}(\phi,\T)
 + {\Phi}^\ast_A {{\cal M}}^A_{1}({\tilde\Phi}) +
 \frac{1}{2} {\Phi}^\ast_A {\Phi}^\ast_B
 {\C{M}}^{BA}_{1}( {\tilde \Phi}) + \, ...,
 \ee
It is important to emphasize that the higher terms such
${{\cal M}}^A_{1}({\tilde \Phi})$
can not be written in a closed form in terms of the finite transformation.

 The full WZ term in the gauge fixed basis is obtained from
 \bref{WZ AI Cl2}
 by means of the canonical transformation \bref{ct to gfb}.
 The antifield independent part will
 have the form
 \be
 {\hat {\cal M}}_{1,\Sigma}(\tilde{\Phi}) ={ \C{M}}_{1}(\phi,\T)+
 (\frac{\delta\Psi({\Phi})}{ \delta {\Phi^A}})
 { {\cal M}}^A_{1}(\tilde\Phi)+...
 \ee
 Note that these  expressions
are not written in terms of the finite transformation.


\section{Example: Non-Abelian Antisymmetric Tensor Field}
\indent

There are numbers of applications of the present formalism.
One is a system of spinning string which has on-shell irreducible
algebra with Weyl and super Weyl anomalies.
Others are N=1 super Yang-Mills theories in d=6,8,10 for which
off-shell formulation has not been known.
The d=10 case may be regarded as an effective theory from superstring.
$W_n$ theories are also supposed to have antifield dependent anomalies.

In this section we will study a system of non-Abelian antisymmetric
tensor field \cite{ft} to show some feature of the reconstruction
procedure.

The classical action is given by
\be
{\cal S}_0= \frac{1}{2} \int\dif^4x \,{\rm tr}\{ A_\mu A^\mu +
B_{\mu\nu}F^{\mu\nu} \}.
\ee
where
$B_{\mu\nu}=-B_{\nu\mu}= B^a_{\mu\nu} T_a$
is the antisymmetric tensor field
and  a vector gauge field $A_{\mu}=A^a_{\mu} T_a$
is playing the role of a lagrange multiplier.
$F_{\mu\nu} = [D_\mu,D_\nu]$ is the field strength with
$D_\mu=\pa_\mu+ A_\mu$.~
$T_a$'s are the generators of some semisimple algebra
 with the algebra $[T_a,T_b]=f^c_{~ab}T_c$.

The classical equation of motions are expressed as
\bea
 \frac{\delta {\cal S}_0}{\delta B_{\mu\nu}}
= F^{\mu\nu},~~~~~~~~~~
 \frac{\delta {\cal S}_0}{\delta A_\mu}
= A_\mu - (D^\A B_{\A\mu}),
\eea
where
$$
D^\A B_{\A\mu}~=~\pa^\A B_{\A\mu}+[A^\A,B_{\A\mu}].
$$
The infinitesimal gauge transformations are
\bea
 \delta_{\Lambda^\B} B_{\mu\nu} = \ep_{\mu\nu\A\B}(D^\A \Lambda^\B),
{}~~~~~~
\delta_{\Lambda^\B} A_\mu = 0
\eea
and the algebra result to be abelian.
This system  has on-shell reducible symmetry, {\it i.e.} the transformation
with the parameter ${\Lambda^\A} = {(D^\A \zeta)}$
is trivial on the classical equation,
\be
\delta_{D^\B \zeta} B_{\mu\nu} = \ep_{\mu\nu\A\B} (D^\A D^\B \zeta)
=\frac{1}{2}\ep_{\mu\nu\A\B}[F^{\A\B },\zeta] \sim 0.
\ee

The infinitesimal transformations are integrated to give finite
transformations and we can write the finite on-shell structure
\bref{actionF}.
The finite gauge transformations are
\bea
\label{FGT}
B'_{\mu\nu} = B_{\mu\nu} + \ep_{\mu\nu\A\B} D^\A \Lambda^\B,
{}~~~~~~~~A'_\mu = A_\mu.
\eea
and the functions characterizing the on shell redundancy in \bref{comp}
are
\be f^{\T^\A}(\Lambda,\zeta,\phi)~=~\Lambda^\A+D^\A~ \zeta, ~~~
\Psi_{\mu\nu,\alpha\beta}=\frac12\epsilon_{\mu\nu\alpha\beta}~\zeta.
\ee
The proper solution of CME in the classical basis is \cite{afsol}
\bea
\nonumber
{\cal S} = \int \dif^4x \, {\rm tr} \{\,\frac{1}{2}(A^\mu A_\mu +
B_{\mu\nu}F^{\mu\nu}) +  \frac{1}{2} B^\ast_{\mu\nu}
\ep^{\mu\nu\A\B} (D_\A c_\B) + \ c^\ast_\A (D^\A \eta)
\\
+ \frac{1}{4} B^\ast_{\mu\nu}~B^\ast_{\A\B}~\ep^{\mu\nu\A\B} \eta
+ {\bar c}^\ast_\A b^\A +\ {\bar \eta}^\ast d +\ \eta'^\ast d' \,\}.
\eea

We go to the GFxB by using the gauge fixing fermion
\be
\label{gauge fermi}
\Psi =  \int \dif^4x \ {\rm tr} \{ \ (\pa_\mu {\bar
c}_\nu)
B^{\mu\nu} + \ (\pa_\mu {\bar \eta}) c^\mu + \ (\pa_\mu {\bar c}^\mu)
\eta' + \ \frac{1}{2} {\bar c}_\A b^\A + \ {\bar \eta}d' \ \},
\ee
giving the gauge fixed action,
\bea
\nonumber
\hat{{\cal S}}^{GFxB} = &&  \int \dif^4x \, {\rm tr} \{\frac{1}{2} A_\mu
A^\mu + \frac{1}{2} B_{\mu\nu} F^{\mu\nu} +
\ \frac12(B^\ast_{\A\B}+\pa_{[\A}{\bar c}_{\B]})
\ep^{\A\B\G\delta} D_\G c_\delta
\\
\nonumber
&&+ \frac{1}{4} (B^\ast_{\A\B}+\pa_{[\A}{\bar c}_{\B]})
              (B^\ast_{\rho\sigma}+\pa_{[\rho}{\bar c}_{\sigma]})
 \ep^{\A\B\rho\sigma} \eta
+ (c^{\ast\G}+\pa^\G{\bar \eta}) D_\G \eta
\\
&&+ (\bar c^{\ast\nu}-\pa_\mu B^{\mu\nu} - \pa^\nu \eta' + \frac{1}{2}b^\nu )
b_\nu + (\bar\eta^{\ast}+d' - \pa_\A c^\A)d +(\eta'^\ast+\pa_\mu\bar c^\mu)d'\
\},
\eea
where $(\pa_{[\A}{\bar c}_{\B]}) \equiv \pa_\A {\bar c}_\B - \pa_\B
{\bar c}_\A$.
The antifield independent part of the proper solution is $S_\Sigma$.

Despite the fact that this model has no true anomaly
if we introduce an algebraic solution of
WZ consistency conditions
it can be used to exemplify our formalism.
We will start our analysis by considering an antifield independent
quantity with ghost number 1 verifying  the WZ consistency condition
$\delta_0 \C{A}_1(B_{\mu\nu},A_\mu,c_\alpha)~\approx~0$;~
\be
{\cal A}_1(B_{\mu\nu},A_\mu,c_\alpha)
 = \frac{1}{2} \int \dif^4x \, \ep^{\A\B\G\delta} \, {\rm tr}
\{B_{\A\B}D_{[\G} c_{\delta]}
\ee
Applying the reconstruction procedure in the ClB \cite{h}\cite{vp}
we determine the antifield dependent anomaly
\be
{\cal A}_1 = \frac{1}{2} \int \dif^4x \, \ep^{\A\B\G\delta} \, {\rm tr}
\{B_{\A\B}D_{[\G} c_{\delta]}
- B^\ast_{\A\B} [B_{\G\delta},\eta]
\}.
\ee
To go to the GFxB we make a canonical transformation \bref{ct to gfb} with
$\Psi$ in \bref{gauge fermi}
\be
\label{ctog}
B^\ast_{\mu\nu} \quad \longrightarrow \quad
B^\ast_{\mu\nu} + \pa_{[\mu}{\bar c}_{\nu]} \, ,
\ee
and find the corresponding anomaly in the GFxB
\bea
\nonumber
\hat{{\cal A}}_1 =
 \frac{1}{2} \int \dif^4x \,\ep^{\mu\nu\A\B}
\, {\rm tr}\{\ B_{\mu\nu}D_{[\A} c_{\B]} - [(B^\ast_{\mu\nu}+
\pa_{[\mu}{\bar c}_{\nu]}),B_{\A\B}] \eta \ \},
\label{GFxAno}
\eea

Now we apply the extended formalism.
We elevate the gauge group parameters $\Lambda^\A$
to the category of fields; $\T^\B(x) = (\T^{\B})^a T_a$
. Their gauge transformations are found by demanding
the weak gauge invariance of (\ref{ocl}). We get using \bref{ocl1}
\be
{\tilde \delta}_{\Lambda^\B, \vep} \T^\A = -\Lambda^\A + D^\A \zeta.
\ee
Observe that the gauge group has increased by the new gauge parameters
$\vep=\vep^a T_a$, but the reducibility is maintained, i.e., for
$\Lambda^\B= D^\B \zeta$~ we have
\be
{\tilde \delta}_{D^\B \zeta, \zeta} \T^\A = 0.
\ee
The fact that it is an off-shell equality makes that the same on-shell
reducibility
as in the original theory is maintained and the extended non-proper
solution is
\bea
{\tilde {\cal S}} = {\cal S} + \int \dif^4x \, {\rm tr}\{ \
\T^\ast_\A(-c^\A + D^\A v) + v^\ast \eta \ \},
\eea
with $v$ the ghosts corresponding to the new gauge parameters $\zeta$.

The antifield independent part of
the local WZ term that cancels the anomaly is using \bref{WZ AI Cl}
\bea
\label{AFIWZT}
{\cal M}_{1}(B_{\mu\nu},A_\mu,\T^\A) && =
-i~\int_0^1 dt~ \int \dif^4x\,{\rm tr}\{
\,\frac{1}{2}\ep^{\mu\nu\rho\sigma}(B^{\mu\nu}+ \ep_{\mu\nu\A\B} D^\A(t\T^\B)
)D_{[\rho} \T_{\sigma]}
\\
&&= \frac{-i}{2} \int \dif^4x \,{\rm tr}\{
\ep^{\mu\nu\rho\sigma} (B_{\mu\nu}+\frac12\ep_{\mu\nu\A\B} D^\A \T^\B)
D_{[\rho} \T_{\sigma]}\},
\eea
and satisfies
\bea
{\tilde \delta}_0 {\cal M}_{1}(B_{\mu\nu},A_\mu,\T^\A)
= ({\cal M}_{1}(B_{\mu\nu},A_\mu,\T^\A) ,{\tilde {\cal
S}})|_{{\tilde \Phi}^\ast =0} \simeq \, i{\cal A}_1
(B_{\mu\nu},A_\mu,c_\alpha) .
\eea

This information is enough to find, from our knowledge of the full
anomaly, the remaining antifield dependent part of the WZ term
which is local. A straightforward
calculation gives
\bea
\nonumber
{\cal M}_1(B_{\mu\nu},A_\mu,\T^\A,v,{B^*}^{\mu\nu})  = &&
{\cal M}_{1}(B_{\mu\nu},A_\mu,\T^\A)  +
\frac{i}{2} \int \dif^4x \,{\rm tr}\{\ep^{\mu\nu\rho\sigma}
(B_{\mu\nu}+\ep_{\mu\nu\A\B}D^\A\T^\B)[B^\ast_{\rho\sigma},v]_+\}
\\
&&-~\frac{i}{8} \int \dif^4x \,{\rm tr}\{\ep^{\mu\nu\rho\sigma}\{ \
\ep_{\mu\nu\A\B}[B^{\ast\A\B},v]_+[B^\ast_{\rho\sigma},v]_+\},
\eea
where the anti commutator $[B^\ast_{\rho\sigma},v]_+$ is understood as
$[B^\ast_{\rho\sigma},v]_+=B^\ast_{\rho\sigma}v+v B^\ast_{\rho\sigma}~$.
Once we have the complete WZ term, we can move to the GFxB.
Making the substitution \bref{ctog}
we obtain the full WZ term
\bea
\nonumber
\hat{{\cal M}}^{(GFxB)}_1 &&= \
-\frac{i}{2} \int \dif^4x \,{\rm tr}\{-B_{\mu\nu}B^{\mu\nu}+
(B^{\mu\nu}+\frac{1}{2}\ep^{\mu\nu\rho\sigma} ( D_{[\rho} \T_{\sigma]}-
[B^\ast_{\rho\sigma}- D_{[\rho} \bar c_{\sigma]},v]_+))\times
\\
\nonumber
&&
(B_{\mu\nu}+\frac{1}{2}\ep_{\mu\nu\A\B} ( D^{[\A} \T^{\B]}-
[B^{\ast\A\B}- D^{[\A} \bar c^{\B]},v]_+))\}.
\label{GFxWZ}
\eea

As we know that this theory is not anomalous we expect that
we can integrate the extra variables and still have a local counter term.
In fact if one makes the
following redefinition of the $\T^\A$ variables
\be
\ep^{\rho\sigma\A\B}D_\A \T_\B \quad \longrightarrow \quad
\ep^{\rho\sigma\A\B}D_\A \T'_\B = \ep^{\rho\sigma\A\B}D_\A \T_\B
+ B^{\rho\sigma} -\frac{1}{2}\ep^{\rho\sigma\A\B}[B^\ast_{\A\B},v]_+,
\ee
one gets the decoupling in the WZ term in the GFxB
\bea
 \hat{{\cal M}}^{(GFxB)}_1 &&= -\frac{i}{2} \int \dif^4x
\,\ep^{\rho\sigma\A\B}\ep_{\A\B\mu\nu}\,{\rm tr}\{(D_\rho \T'_\sigma)
(D^\mu\T'^\nu)\} + \frac{ia}{2} \int \dif^4x \,{\rm tr}\{
B_{\mu\nu} B^{\mu\nu}\}
\\
\nonumber
&& \equiv {\cal N}_1(\T'_\A) + {\cal O}_1(B_{\mu\nu}).
\eea
We can verify that $\C{N}_1$ is BRST invariant and see that
\be
{\cal O}_1 = \frac{ia}{2} \int \dif^4x \,\rm{tr}\{\,
B_{\mu\nu}B^{\mu\nu} \,\}
\ee
cancels exactly the complete anomaly in GFxB,
\bea
\nonumber
({\cal O}_1,{\tilde {\cal S}})  = \frac{i}{2} \int \dif^4x
\,\ep^{\A\B\G\delta}\,\rm{tr}\{\ B_{\A\B}D_{[\G} c_{\delta]} -
[B^\ast_{\A\B},B_{\G\delta}]\eta \}
 = i\,\hat{{\cal A}}_1.
\eea
This ends our analysis of the antisymmetric tensor field.

 \section{Conclusions}
 \indent

 We have constructed the one loop effective action for  general gauge
theories in a PV regularization scheme.  The non-invariance of the
effective action gives the anomaly.  The anomaly is parametrized in
terms of the PV regulator and the BRST transformation. This
parametrization is very useful in order to study the antifield dependence.
The anomaly depends on fields and antifields in a non-trivial way  for
theories with an on-shell structure, {\it i.e.} theories with an open
algebra or reducible on-shell algebra.

Introducing  extra degrees of freedom corresponding to the classical
gauge degrees of freedom they become dynamical at quantum level.
We have constructed the local WZ term which is depending in general
on the antifields.  The antifield independent part, in the classical
basis, has the usual form in terms of the anomaly and the finite gauge
transformations. The full WZ term in the classical basis is obtained
using the Afs perturbation methods. The WZ term in gauge fixed basis
is obtained by the straightforward canonical transformation.

In this paper we did not discuss the issue of quantization of the extra
variables, which will be discussed in a future work where we will also study
in detail the finite form of the gauge structure and the corresponding
 infinitesimal structure functions.
\vskip 6mm

{\bf Acknowledgments}
We acknowledge discussions with Profs. T. Kugo, N. Nakanishi, I. Ojima,
W. Troost, M. Henneaux and a careful reading of the manuscript to Prof.
T. Kugo and Dr. J. Par\'\i s. This work has
been partially supported by CYCYT under contract number AEN93-0695,  by
Comissionat per Universitats i Recerca de la Generalitat de Catalunya and
by Commission of European Communities contract CHRX-CT93-0362(04).

\appendix
\section{  On-shell Quasigroup Structure }
 \indent

 Let us consider the action of an on-shell quasigroup G which is locally
 described by a set of  parameters $\theta^\alpha$
 on a manifold ${\cal M}$ parametrized by the classical fields ${\phi^i}$
 \bea
 \nonumber
 & F: {\cal M} \times {\cal G}\rightarrow {\cal M}
 \\
 \label{actionF}
 &\quad\quad\quad\quad(\phi^i,~\theta^\alpha) \mapsto F^i (\phi,\theta)
 \eea
 with the following properties \footnote{Here we only write the relevant
properties of the on-shell structure useful for the discussion below.}.
\vskip 4mm

i) \quad Invariance of the action;
\be
 {\cal S}_{0}( F (\phi,\theta))={\cal S}_{0}(\phi).
\ee
As a consequence we have the following relation between the equations of
motion
\be
\label{eqmot}
S_{0,j}(F(\phi,\T))~=~\tilde S_j^{~k}(\phi,\T)S_{0,k}(\phi).
\ee
\vskip 4mm

 ii) \quad On-shell redundancy in the  parametrization;
 \be
 \label{comp}
 F^i(\phi, f(\theta,\epsilon;\phi)) =
 F^i(\phi,\theta) + \Psi^{ij}(\theta,\epsilon;\phi) {\cal S}_{0,j}(\phi)
 \label{Psi}
 \ee
 where $f(\theta,\epsilon;\phi)$ and $\Psi^{ij}(\theta,\epsilon;\phi)$
 represents the on-shell reducibility.
\vskip 4mm

 iii) \quad Composition law;
 \be
 F^i (F (\phi,\theta),\theta')=
 F^i (\phi,\varphi(\theta,\theta';\phi)) +
 M^{ij}(\theta,\theta';\phi) {\cal S}_{0,j}(\phi),
 \label{openF}
  \ee
 where $\varphi^\A(\theta,\theta';\phi)$ represents the composition
 function of the  parameters of quasigroup G and
 $M^{ij}(\theta,\theta';\phi)$ represents the
 open character of the finite gauge transformations.
\vskip 4mm

iv) \quad Associativity law;
\be
\label{malex}
\varphi^\alpha \left(\varphi\left(\theta,\theta';\phi\right),\theta''
;\phi\right) = f^\alpha
\left(\varphi\left(\theta,\varphi\left(\theta',\theta'';F(\phi,\T)
\right);\phi\right), \eta\left(\theta,\theta',\theta'';\phi\right)
,\phi\right)
\\ + M^{\alpha i}(\T,\T',\T'';\phi){\cal S}_{0,i}(\phi),
\ee
where $M^{\alpha i}(\theta,\theta',\theta'';\phi)$ and
$\eta\left(\theta,\theta',\theta'';\phi\right)$
represent the modified on-shell associativity law for the parameters.
\vskip 4mm

v) \quad On-shell structure of the on-shell functions $M^{ij}$
\bea
\nonumber
&&
\{M^{ij}\left(\T,\varphi(\T',\T'';F(\phi,\T));\phi\right)
+~M^{ik}(\T',\T'';F(\phi,\T))\tilde S_k^{~j}(\phi,\T)~\}
\\
\nonumber
&&~-~\{
 M^{ij}(\varphi(\T,\T',\phi),\T'';\phi)~+~
\frac{\pa F^i(F(\phi,\varphi(\T,\T';\phi)),\T'')}{\pa F^k}
M^{kj}(\T,\T';\phi)~+~
\\
\nonumber
&&
\frac{\pa F^i\left(\phi,f(\varphi(\T,\varphi(\T',\T'';F(\phi,\T));\phi)),
\eta(\T,\T',\T'';\phi);\phi\right)}{\pa f^\A}
M^{\A j}(\T,\T',\T'';\phi)~+~
\\
&&
\nonumber
+~\Psi^{ij}\left(\varphi(\T,\varphi(\T',\T'';F(\phi,\T));\phi),
\eta(\T,\T',\T'';\phi);\phi \right)~\}~
+~\frac{\partial F^k(\phi,\T)}{\pa \T^\A}
M^{\A i}(\T,\T',\T'';\phi)
\\
\label{e43'}
&&~=~
-~M^{ijk}(\T,\T',\T'';\phi)S_{0,k}(\phi),
\eea
where $ M^{ijk}(\T,\T',\T'',\phi)$ represents the non-closure of the
on-shell structure.

 In a  general situation the functions $M^{ijk}$ are not close on-shell.
New structure functions appear when we perform the
composition of four or more transformations.

The structure functions appearing in the solution of the classical master
equation are directly related to these functions
{\it only in the classical basis}. For example those appearing in the
algebra \bref{openRR} and \bref{red-off} are given as
 $$
 {R^i}_{\A}( \phi)~:=~ \left.\frac{\partial F^i(\phi,
 \T)}{\partial\T^\A}\right|_{\T= 0},~~~~~
 T^{\gamma}_{\alpha \beta}(\phi) :=-
 \left(\frac{\partial^2
 \varphi^{\gamma}(\theta,\theta';\phi)}{\partial\theta^\alpha
 \partial\theta'^\beta} - (\B \leftrightarrow \A)
 \right)_{\theta=\theta'=0},
 $$
 $$
 E^{ij}_{\A \B}(\phi)  := -
 \left(\frac{\partial
 M^{ij}(\theta,\theta';\phi)}{\partial\T^\A \T'^\B}
 - (\B\leftrightarrow\A) \right)_{\T=\T'=0},~~~~~
 $$
 \be
 Z^\A_a(\phi) := \left. \frac{\partial f^\A (\T,\ep,\phi)}
 {\partial \ep^a}\right|_{\T=\ep=0},~~~~~
 V^{ij}_a(\phi) := - \left.\frac{\partial \Psi^{ij}(\T,\ep,\phi)}
 {\partial \ep^a}\right|_{\T=\ep=0}
 \ee
The ones appearing in the generalized Jacobi identity \bref{4-4}
are :
\be
F^a_{\A\B\G}(\phi) :=~ -\frac13~\sum_{P\in{\rm Perm}[\A\B\G]} (-1)^P
\left(\frac{\partial^3\eta^a(\T,\T',\T'',\phi)}{\partial\T^
\A\partial\T'^\B\partial\T''^\G}
\right)_{\T=\T'=\T''=0}
\ee
and
\be
D^{\nu i}_{\A\B\G}(\phi) :=~ \frac13~\sum_{P\in{\rm Perm}[\A\B\G]} (-1)^P
\left(\frac{\partial^3
M^{\nu i}(\T,\T',\T'',\phi)
}{\partial\T^
\A\partial\T'^\B\partial\T''^\G}
\right)_{\T=\T'=\T''=0}.
\ee

 A detail analysis of the on-shell structure and the relation with the
work of \cite{bv84} will be published elsewhere.

\vskip 5mm

 From the on-shell composition law \bref{openF} we can obtain the
 on-shell Lie equation by multiplying an operator
 $\left. \frac{\partial}{\partial{\T'}}
 \right|_{\T'=0}$ on it. It is explicitly
\be
 \frac{\partial F^i(\phi,\T)}{\partial\T^\A} = R^i_\B(F(\phi,\T))
 \lambda^\B_{~\A}(\T,\phi) - \lambda^\B_{~\A}(\T,\phi)
 \left.\frac{\pa M^{ij}(\T,\T';\phi)}{\pa \T'^\B}\right|_{ \T'=0}
 {\cal S}_{0,j}(\phi) \, ,
 \label{eqtr}
 \ee
 where $\lambda^\B_{~\A}(\T,\phi)$ is the inverse matrix of
 \be
 \mu^\A_{~\B}(\T,\phi)= \left.\frac{\partial
\vp^\A(\T,\T',\phi)}{\partial\T'^\B}
 \right|_{ \T'=0} \, .
 \label{mu}
 \ee
 On the other hand if we operate  $\left. \frac{\partial}{\partial{\T}}
 \right|_{\T=0}$ on \bref{openF} we have
\be
 \frac{\partial F^i(\phi,\T)}{\partial \phi^k}~R^k_\A(\phi)~=~
 \frac{\partial F^i(\phi,\T)}{\partial \T^\B}~\tilde \mu^\B_{~\A}
(\T,\phi)~+~
 \left.\frac{\pa M^{ij}(\T',\T;\phi)}{\pa \T'^\B}\right|_{ \T'=0}
 {\cal S}_{0,j}(\phi) \, ,
 \label{eqtrB}
 \ee
 where
  \be
 \label{vecZA}
  \tilde \mu^\B_{~\A}(\T,\phi) :=\left.\frac{\partial\vp^\B(\T',\T;\phi)}
 {\partial\T'^\A}\right|_{\T'=0}.
 \ee

Finally applying operator  $\left. \frac{\partial}{\partial{\ep}}
 \right|_{\ep=0}$ on \bref{Psi} we get
\be
 \frac{\partial F^i(\phi,\T)}{\partial \T^\B}~Z^\B_a(\T,\phi)~=~
 \left.\frac{\pa \Psi^{ij}(\T,\ep;\phi)}{\pa \ep^a}\right|_{ \ep=0}
 {\cal S}_{0,j}(\phi) \, ,
 \label{eqtrC}
 \ee
 where
 \be
 \label{vecZB}
 Z^{\alpha}_{a}(\T, \phi) := \left. \frac{\partial f^\A (\T,\ep,\phi)}
 {\partial \ep^a}\right|_{\ep = 0}.
 \ee

\end{document}